\documentclass[10pt]{article}

\usepackage{amssymb}
\usepackage{amsmath}

\begin{document}

\begin{center}
\Large{\bf THE SCHR\"{O}DINGER PICTURE AND THE ZERO-POINT RADIATION}\\[0.3cm]
\end{center}

\vspace{0.5cm}

\begin{center}
A. J. FARIA, H. M. FRAN\c{C}A\footnote{e-mail: hfranca@if.usp.br},
C. P. MALTA and R. C. SPONCHIADO \\[0.3cm]
\end{center}

\begin{center}
{\em Instituto de F\'{i}sica, Universidade de S\~{a}o Paulo \\
C.P. 66318, 05315-970 S\~{a}o Paulo, SP, Brazil}
\end{center}

\vspace{0.5cm}

\begin{abstract}
Dalibard, Dupont-Roc and Cohen-Tannoudji (J. Physique
43 (1982) 1617; 45 (1984) 637) used the
Heisenberg picture to show that the atomic transitions, and the
stability of the ground state, can only be explained by introducing
radiation reaction and vacuum fluctuation forces. Here we consider the
simple case of nonrelativistic charged harmonic oscillator, in one
dimension, to investigate how to take into account the radiation
reaction and vacuum fluctuation forces within the Schr\"{o}dinger
picture. We consider classical vacuum fields and large mass
oscillator. 
\end{abstract}


\vspace{2cm}

We start by indicating the importance of the radiation reaction
and the vacuum fluctuation forces to the understanding of the atomic
transitions, and the atomic stability, using the Heisenberg
picture. Consider a physical system like the hydrogen atom. Its
Hamiltonian is
\begin{equation}
H_S = \frac{\vec{p\,}^2}{2m} - \frac{e^2}{r}
\label{1}
\end{equation}
and the atomic states are such that
\begin{equation}
H_S|{\rm{vac}},a\rangle=\epsilon_a|{\rm{vac}},a\rangle ,
\label{2}
\end{equation}
where $\epsilon_a$ is the energy and $|{\rm{vac}},a\rangle \equiv |{\rm{vac}}
\rangle |a\rangle$ denotes the state in which the atom is in the stationary state
  $|a\rangle$, and the field is in its vacuum state $|{\rm{vac}}\rangle$ of no photons.

Dalibard, Dupont-Roc and Cohen-Tannoudji \cite{Dalibard} have discussed
the role of the vacuum fluctuations and the radiation
reaction forces, with the identification of their respective
contributions, in the domain of the atomic transitions with
emission of electromagnetic radiation. For clarity's sake we
summarize their main conclusion.

Using a perturbative calculation based on the Heisenberg
picture, Dalibard et al. \cite{Dalibard} concluded that the
variation with time of the energy of the system is such that
\begin{eqnarray}
\langle {\rm{vac}},a|\frac{dH_S}{d\,t}|{\rm{vac}},a\rangle= - \frac{2}{3}
\frac{e^2}{c^3}\langle a| (\ddot{\vec{r}})^2 |a\rangle + 
\nonumber \\ +
\frac{2}{3}\frac{e^2}{c^3}
\left[ \sum_{b\,(\epsilon_{b}>\epsilon_{a})}\langle a| \ddot{\vec{r}}
|b\rangle \cdot \langle b| \ddot{\vec{r}} |a\rangle -
\sum_{b\,(\epsilon_{b}<\epsilon_{a})} \langle a| \ddot{\vec{r}}
|b\rangle \cdot \langle b| \ddot{\vec{r}} |a\rangle \right] .
\label{3}
\end{eqnarray}
The first term in (\ref{3}) is the contribution of radiation reaction
whereas the second, and the third terms, are the contributions of
the vacuum fluctuation forces.

It is straightforward to show that (\ref{3}) can be written as
\begin{equation}
\langle {\rm{vac}},a|\frac{dH_S}{d\,t}|{\rm{vac}},a\rangle= - \frac{4}{3}
\frac{e^2}{c^3} \sum_{b(\epsilon_b<\epsilon_a)} \langle a|
\ddot{\vec{r}} |b\rangle \cdot \langle b| \ddot{\vec{r}} |a\rangle .
\label{4}
\end{equation}

We note that, ``if self reaction was alone'' (see the first term in
(\ref{3})), ``the atomic ground state would not be stable, since the
square of the acceleration has a non zero average value in such a
state'' \cite{Dalibard}. Moreover, ``such a result is extremely
simple and exactly coincides with what is found in classical
radiation theory'' \cite{Dalibard}.

The complete result (see equation (\ref{4})), which includes the vacuum
forces, is even more satisfactory because ``the electron in the
vacuum can only lose energy by cascading downwards to lower
energy levels. In particular, the ground state is stable since it
is the lowest state. The ground state cannot be stable in the absence
of vacuum fluctuations which exactly balance the energy loss due
to self reaction. In other words, if self reaction was alone, the
ground state would collapse and the atomic commutation relation
$ [x,p_x]=i\hbar $ would not hold'' \cite{Dalibard}.

As stated in reference \cite{Dalibard}, ``all self reaction
effects, which are independent of $\hbar$, are strictly identical to
those derived from classical radiation theory. All vacuum
fluctuation effects, which are proportional to $\hbar$ can be interpreted
by considering the vibration of the electron induced by a random
field having a spectral power density equal to $\hbar\omega/2$
per mode''. Therefore, in several situations, the vacuum field can be
successfully replaced by a {\em classical} random field
\cite{Boyer,Milonni76}, so that in the vacuum state the
electric and magnetic fields can be considered as fluctuating about
their mean zero value.

The classical vacuum electric field to be considered here is the random vacuum
electric field of Stochastic Electrodynamic (SED). Following the
notation of Boyer \cite{Boyer}, one can write
\begin{equation}
E_0(t) = \sum_{\lambda=1}^{2} \int d^3k
\epsilon_x(\vec{k},\lambda) \frac{\sqrt{\hbar\omega/2}}{2\pi}
\left[ e^{i\theta(\vec{k},\lambda)} e^{-i\omega t} +
e^{-i\theta(\vec{k},\lambda)} e^{i\omega t} \right] ,
\label{6b}
\end{equation}
where $\theta(\vec{k},\lambda)$ are random phases statistically
independent and uniformly distributed in the interval $[0,2\pi]$,
$\vec{k}$ is the wave vector such that $|\vec{k}| = \omega/c$, and
$\epsilon_x(\vec{k},\lambda)$ is the polarization vector projected in
the $x$ axis, with $\lambda = 1 \, , \, 2$.

The {\em atomic stability} and the {\em atomic transitions} can only
be understood by the introduction of the radiation reaction and the
vacuum fluctuation forces. Without any of these forces, an excited
state cannot decay or the ground state is unstable
\cite{Dalibard,Franca}. Therefore, we shall consider these
electromagnetic forces within the Schr\"{o}dinger
picture. The system we shall study is a charged harmonic oscillator
with natural frequency $\omega_0$ and mass $m$, already considered in
a previous work \cite{faria}.

By considering the dipole (or long wavelength) approximation the
onedimensional Schr\"{o}dinger equation takes the form
\begin{equation}
i \hbar \frac{\partial \psi(x,t)}{\partial\, t} = \left[
\frac{1}{2m} \left( -i\hbar \frac{\partial}{\partial x} -
\frac{e}{c}A_x(t) \right)^2 + \frac{m \omega_{0}^{2} x^2}{2}
\right] \psi(x,t) ,
\label{14}
\end{equation}
where $A_x(t)$ is $x$ component of the vector potential acting on the
charged particle. At this point the exact analytical form of $A_x(t)$
is not known. For the moment we shall simply assume that $A_x(t)$ is a
c-number that varies with $t$ and is {\em independent} of $x$. It
should be noticed that this assumption is valid provided that $mc^2
\gg \hbar \omega_0$.

The time independent Schr\"{o}dinger equation has a ground state
solution $\phi_0(x)$ such that
\begin{equation}
\phi_0(x) = \left( \frac{m \omega_0}{\pi \hbar}
\right)^{\frac{1}{4}} \exp{\left( - \frac{m \omega_0 x^2}{2 \hbar}
\right)} .
\label{15}
\end{equation}

Moreover, we see that
\begin{equation}
\int_{-\infty}^{\infty} dx\, \phi_{0}^{2}(x)
x^2 = \frac{\hbar}{2 m \omega_0} .
\label{16}
\end{equation}

The time dependent equation (\ref{14}) has an exact solution that
can be written as
\begin{equation}
\psi(x,t) = \phi_{0} \left( x-q_c(t) \right) \exp{ \left\{
\frac{i}{\hbar} \left[ \left( p_c(t) +\frac{e}{c}A_x(t) \right) x - g(t)
\right] \right\} } ,
\label{17}
\end{equation}
where the functions $q_c(t)$, $p_c(t)$ and $g(t)$ are unknown
c-numbers that will be determined by the substitution of (\ref{17})
into (\ref{14}). This is an old procedure, introduced by
Schr\"{o}dinger (1926) in a famous paper entitled ``The Continuous
Transition from Micro to Macro-Mechanics'' (see reference
\cite{Schrodinger}, pg. 41). With
the above substitution, and requiring that both the real and the imaginary
parts of the resulting Schr\"{o}dinger equation must be zero for
arbitrary values of $x$ and $t$, we get the following equations:
\begin{equation}
p_c(t) = m \dot{q}_c(t) ,
\label{18}
\end{equation}
and
\begin{equation}
\dot{p}_c(t) = - m\omega_{0}^2 q_c(t) - \frac{e}{c}
\frac{\partial A_x(t)}{\partial t} .
\label{19}
\end{equation}
We also obtain the equation $2\dot{g}(t) = \hbar\omega_0 +
m\dot{q}_c^2(t) - m\omega_0^2 q_c^2(t)$, which can be written as
\begin{equation}
g(t) = \frac{\hbar\omega_0 t}{2} + \frac{m}{2}
\int_0^t dt' \left( \dot{q}_c^2(t') - \omega_0^2 q_c^2(t') \right) .
\label{20}
\end{equation}
One can combine (\ref{18}) and (\ref{19}) obtaining the differential
equation
\begin{equation}
m \ddot{q}_c(t) = -m\omega_0^2 q_c(t) + e E_x(t) ,
\label{21}
\end{equation}
where we have used the fact that $cE_x(t) = -\partial A_x(t)/\partial
t$. Notice that, by assumption, every term in (\ref{21}) is a
c-number. According to our assumption, we shall write
\begin{equation}
E_x(t) = E_0(t) + E_{RR}(t) ,
\label{22}
\end{equation}
where $E_0(t)$ (see equation (\ref{6b}))
is the classical vacuum field and $E_{RR}(t)$ is the classical
radiation reaction field (the particle is charged therefore the
radiation reaction field must contribute to $E_x(t)$).

The correct expression for the classical radiation reaction force
$eE_{RR}(t)$ is more difficult to obtain because, according to the
Schr\"{o}dinger picture, the charged particle does not have a precise
location. One can only say that
\begin{equation}
|\psi(x,t)|^2 =
\left( \frac{m\omega_0}{\pi\hbar} \right)^{\frac{1}{2}}
\exp{\left[ - \frac{m \omega_0 (x - q_c(t))^2}{\hbar} \right]} ,
\label{23}
\end{equation}
is the probability density. Notice that, in order to obtain
(\ref{23}), one must solve (\ref{21}) which depends on the still
undefined radiation reaction force $eE_{RR}(t)$. This force, however,
can be precisely defined in the case of large mass, so that
$mc^2 \gg \hbar\omega_0$. In this case, one can safely
consider that
\begin{equation}
eE_{RR}(t) = \frac{2e^2}{3c^3} \stackrel{...}{q}_c(t)
\label{24}
\end{equation}
is a good approximation because the Gaussian (\ref{23}) is so
narrow that the harmonically bound particle has a
trajectory. Based on these considerations we conclude that the
expression (\ref{24}) is valid in the case $mc^2 \gg \hbar \omega_0$,
which is consistent with the {\em long wavelength}
approximation. Therefore the equation (\ref{21}) can be written as 
\begin{equation}
\ddot{q}_c(t) + \omega_0^2 q_c(t) \simeq \frac{e}{m} E_0(t) +
\frac{2e^2}{3mc^3} \stackrel{...}{q}_c(t) ,
\label{25}
\end{equation}
where $E_0(t)$ is given by (\ref{6b}). The stationary solution of this
equation is
\begin{equation}
q_c(t) = \frac{e}{m} \sum_{\lambda=1}^{2} \int d^3k
\epsilon_x(\vec{k},\lambda) \frac{\sqrt{\hbar\omega/2}}{2\pi}
\left[ \frac{e^{i\theta(\vec{k},\lambda)} e^{-i\omega t} +
e^{-i\theta(\vec{k},\lambda)} e^{i\omega t}}{\omega_0^2 - \omega^2 -
i\frac{2e^2}{3mc^3} \omega^3} \right] .
\label{stat}
\end{equation}

Using the {\em stationary} solution (\ref{stat}), and the wave
function (\ref{17}), we can calculate the mean square value of the
particle position. This quantity is obtained by averaging over the
random phases present in (\ref{stat}). The expectation value of the
operator $x^2$ is
\begin{equation}
\overline{x^2(t)} =
\int_{-\infty}^{\infty}dx |\psi(x,t)|^2\, x^2 ,
\label{26}
\end{equation}
and taking into account the expressions (\ref{17}) and (\ref{15}), one
can show that
\begin{eqnarray}
\overline{x^2(t)} & = &
\int_{-\infty}^{\infty}dx \phi_{0}^2(x-q_c(t))\,
\left[\left(x-q_c(t)\right)^2+q_c^2(t) \right] 
= \nonumber \\ & = &
\frac{\hbar}{2m\omega_0} + q_c^2(t) .
\label{27}
\end{eqnarray}

The average over the random variables (indicated by the symbol
$\langle \, \rangle$) is such that \cite{Boyer}
\begin{equation}
\begin{array}{c}
\langle e^{i\theta(\vec{k},\lambda)} e^{i\theta(\vec{k}',\lambda')} \rangle =
\langle e^{-i\theta(\vec{k},\lambda)} e^{-i\theta(\vec{k}',\lambda')}
\rangle = 0 ,
\\
\langle e^{i\theta(\vec{k},\lambda)} e^{-i\theta(\vec{k}',\lambda')} \rangle =
\delta_{\lambda\lambda'} \delta^3(\vec{k} - \vec{k}') .
\end{array}
\label{ran}
\end{equation}
Hence, applying the random average to the expression (\ref{27}), we
obtain
\begin{equation}
\langle \overline{x^2} \rangle = \frac{\hbar}{2m \omega_0} +
\langle q_c^2(t) \rangle .
\label{28}
\end{equation} 
Notice that $\langle \overline{x} \rangle =
\langle q_c(t) \rangle = 0$ in the stationary regimen.

Using the stationary solution (\ref{stat}), the average of $q_c^2(t)$ over the
random phases is
\begin{equation}
\langle q_c^2(t) \rangle =
\frac{2 e^2}{3\pi m^2c^3} \int_0^{\infty}
d\omega \frac{\hbar\omega^3}{ ( \omega^2 - \omega_0^2 )^2 +
\left( \frac{2e^2}{3mc^3} \right)^2 \omega^6} \simeq \frac{\hbar}{2m\omega_0} .
\label{28b}
\end{equation}
The integration in equation (\ref{28b}) was performed assuming
$\frac{2e^2 \omega_0}{3mc^3} \ll 1$. Notice that the result (\ref{28b}) is
charge independent, however, it is not valid for $e = 0$ as an
uncharged mass does not couple to the electromagnetic field, so in this
case we have a free oscillator (see equation (\ref{14})). We must
recall that the function $q_c(t)$ is only a parameter, present in the
solution (\ref{17}) of the Schr\"{o}dinger equation (\ref{14}) only if
$e \neq 0$ \cite{faria,nikolic}.

Substituting the result (\ref{28b}) in the expression (\ref{28}), we get
\begin{equation}
\langle \overline{x^2} \rangle  = 
\frac{\hbar}{m \omega_0} ,
\label{29}
\end{equation}
corresponding to a ground state energy that is twice the correct
value.

One can understand the discrepancy in the following manner. The study
of the emission of radiation, by charged oscillators and atomic
systems, is usually made by performing {\em perturbative} calculations
that involves the vacuum fields and the radiation reaction fields
\cite{Merzbacher,Davydov}. The calculation presented here, though
semiclassical, is {\em nonperturbative}. Therefore the Schr\"{o}dinger
picture should lead to the correct result as obtained within the
Heisenberg picture. Nevertheless we have obtained a discrepancy by a
factor of 2. The inclusion of the zero-point fluctuations within the
Schr\"{o}dinger picture is problematic, indicating that it is not
equivalent to the Heisenberg picture.

As far as we know, Dirac was the first physicist to call
the attention to the fact that the Schr\"{o}dinger and the Heisenberg
pictures are not equivalent \cite{Dirac}. According to him, people usually say that
the Heisenberg and the Schr\"{o}dinger pictures of quantum
mechanics are equivalent and that one can use whichever one likes
indiscriminately. 
The examples given by Dirac to show the lack of equivalence
of the Heisenberg and the Schr\"{o}dinger pictures were the Lamb shift
and the electron anomalous magnetic moment.
Therefore, since these effects involve quite complicated calculations,
that result in very small corrections to the atomic energy
levels, the physicists community did not give much attention to Dirac's
discovery. We have shown here that this lack of equivalence is also
exhibited by a {\em simple} nonrelativistic system, namely the charged
harmonic oscillator.

\section*{Acknowledgements}
We thank Professor Jean-Pierre Vigier for many comments concerning
this subject.
We acknowledge the financial support from Funda\c c\~ao de Amparo \`a
Pesquisa do Estado de S\~ao Paulo (FAPESP) and Conselho Nacional de
Desenvolvimento Cient\'{\i}fico e Tecnol\'ogico (CNPq Brazil).

\end{document}